\documentclass[journal]{IEEEtran}
\IEEEoverridecommandlockouts
\usepackage{cite}
\usepackage{amsmath,amssymb,amsfonts}
\usepackage{graphicx}
\usepackage{textcomp}
\usepackage{xcolor}
\usepackage{multirow}
\usepackage{makecell}
\usepackage{algorithm,algcompatible}

\def\BibTeX{{\rm B\kern-.05em{\sc i\kern-.025em b}\kern-.08em
    T\kern-.1667em\lower.7ex\hbox{E}\kern-.125emX}}

\long\def\comment#1{}

\newfont{\bbb}{msbm10 scaled 700}

\newfont{\bb}{msbm10 scaled 1100}
\newcommand{\CC}{\mbox{\bb C}}

\newcommand{\RR}{\mbox{\bb R}}

\newcommand{\EE}{\mbox{\bb E}}

\newcommand{\av}{{\bf a}}

\newcommand{\ev}{{\bf e}}

\newcommand{\nv}{{\bf n}}

\newcommand{\sv}{{\bf s}}

\newcommand{\vv}{{\bf v}}
\newcommand{\xv}{{\bf x}}
\newcommand{\yv}{{\bf y}}

\newcommand{\Am}{{\bf A}}

\newcommand{\Fm}{{\bf F}}
\newcommand{\Gm}{{\bf G}}
\newcommand{\Hm}{{\bf H}}
\newcommand{\Id}{{\bf I}}

\newcommand{\Pm}{{\bf P}}

\newcommand{\Rm}{{\bf R}}

\newcommand{\Um}{{\bf U}}
\newcommand{\Wm}{{\bf W}}
\newcommand{\Vm}{{\bf V}}

\newcommand{\Pc}{{\cal P}}

\newcommand{\Lambdam}{\hbox{\boldmath$\Lambda$}}

\newcommand{\Sigmam}{\hbox{\boldmath$\Sigma$}}

\newcommand{\Thetam}{\hbox{\boldmath$\Theta$}}

\newcommand{\eqdef}{\stackrel{\Delta}{=}}

\newcommand{\herm}{{\sf H}}

\newcommand{\transp}{{\sf T}}

\DeclareMathOperator*{\argmax}{arg\,max}

\newtheorem{remark}{Remark}

\newtheorem{lemma}{Lemma}
\newtheorem{theorem}{Theorem}

\newtheorem{definition}{Definition}

\begin{document}

\title{Asymptotically Near-Optimal Hybrid Beamforming for mmWave IRS-Aided MIMO Systems

\author{Jeongjae~Lee,~\IEEEmembership{Student Member,~IEEE}
        and~Songnam~Hong,~\IEEEmembership{Member,~IEEE}
\thanks{J. Lee and S. Hong are with the Department of Electronic Engineering, Hanyang University, Seoul, Korea (e-mail: \{lyjcje7466, snhong\}@hanyang.ac.kr).}

\thanks{This work was supported in part by the Technology Innovation Program (1415178807, Development of Industrial Intelligent Technology for Manufacturing, Process, and Logistics) funded By the Ministry of Trade, Industry \& Energy(MOTIE, Korea) and in part by the Institute of Information \& communications Technology Planning \& Evaluation (IITP) under the artificial intelligence semiconductor support program to nurture the best talents (IITP-(2024)-RS-2023-00253914) grant funded by the Korea government(MSIT).}
}
}

\maketitle

\begin{abstract}
Hybrid beamforming is an emerging technology for massive multiple-input multiple-output (MIMO) systems due to the advantages of lower complexity, cost, and power consumption. Recently, intelligent reflection surface (IRS) has been proposed as the cost-effective technique for robust millimeter-wave (mmWave) MIMO systems. Thus, it is required to jointly optimize a reflection vector and hybrid beamforming matrices for IRS-aided mmWave MIMO systems. Due to the lack of RF chain in the IRS, it is unavailable to acquire the TX-IRS and IRS-RX channels separately. Instead, there are efficient methods to estimate the so-called effective (or cascaded) channel in literature. We for the first time derive the near-optimal solution of the aforementioned joint optimization only using the effective channel. Based on our theoretical analysis, we develop the practical reflection vector and hybrid beamforming matrices by projecting the asymptotic solution into the modulus constraint. Via simulations, it is demonstrated that the proposed construction can outperform the state-of-the-art (SOTA) method, where the latter even requires the knowledge of the TX-IRS and IRS-RX channels separately. Furthermore, our construction can provide robustness for channel estimation errors, which is inevitable for practical massive MIMO systems.
\end{abstract}


\section{Introduction}\label{sec:Intro}
Hybrid beamforming is a promising technique for massive multiple-input multiple-output (MIMO) systems due to the advantages of lower complexity, cost, and power consumption \cite{molisch2017hybrid, ahmed2018survey, di2020hybrid}. This approach uses the combination of analog beamformers in the RF domain, together with digital beamforming in the baseband, connected to the RF with a smaller number of up/down-conversion chains (i.e., RF chains). This hybrid structure is motivated by the fact that the number of RF chains is only lower-limited by the number of data streams, whereas the beamforming gain is attained by the number of antenna elements if proper analog beamforming is constructed. There have been many efforts to optimize the hybrid beamforming \cite{el2012capacity, el2014spatially, Heath2016, Yu2016, lin2019hybrid}. Unfortunately, it is quite demanding due to the non-convexity which is caused by the modulus constraints on the analog precoder and combiner. In single-user MIMO (SU-MIMO) systems, the asymptotically optimal solution was derived in \cite{el2012capacity}, wherein the analog beamforming focused on the array gains to a limited number of multiple paths in the RF domain, while the digital beamforming focused on the multiplexing data streams and the power-allocation in the baseband. Beyond the asymptotic results, a number of practical constructions of the hybrid beamforming have been proposed in \cite{el2014spatially, Heath2016,  Yu2016} for millimeter-wave (mmWave) SU-MIMO systems. Exploiting the sparsity of mmWave channels, the optimization problem of the hybrid beamforming was recast as a sparsity-constrained matrix reconstruction problem \cite{el2014spatially}. Then, it can be efficiently solved via compressed-sensing algorithms (e.g., orthogonal matching pursuit (OMP)).  In \cite{Heath2016}, the analog beamformers were optimized by projecting the optimal fully-digital beamformers into the modulus constraints, where the digital beamformers are simply derived via singular value decomposition (SVD). Also, the manifold optimization (MO)-based method was proposed in \cite{Yu2016}, where the non-convex modulus constraints are tackled via the Riemannian submanifold.

Recently, intelligent reflecting surface (IRS) (a.k.a. reflecting intelligent surface (RIS)) has been proposed as the cost-effective technique for robust mmWave MIMO systems \cite{di2020smart, pei2021ris}. An IRS consists of a uniform array with a massive number of reflective elements, each of which can control the phase and the reflection angle of an incident signal so that the received power of the intended signal is improved. Thus, IRS-aided hybrid beamforming can be a good candidate to enhance the robustness and the spectral efficiency of mmWave MIMO systems \cite{wu2019towards}. Toward this, it is necessary to jointly optimize the reflection vector and the hybrid beamforming, which becomes more challenging than the aforementioned optimization of the hybrid beamforming only. In \cite{Wang2021}, the reflection vector and the hybrid beamforming were optimized using the MO-based two-state algorithm. However, the joint optimization was not considered, although they influence each other. By overcoming this limitation, the hybrid beamforming proposed in \cite{9743307} can achieve a near-optimal spectral efficiency in the large-system limit. Despite its superior asymptotic-performance, there is still a room to enhance the performance in practical (or large but finite size) systems. In addition, to construct the reflection vector in \cite{9743307}, it requires the channel state information (CSI) of the TX-IRS and IRS-RX channels (i.e., $\Hm_{\rm TI}$ and $\Hm_{\rm IR}$ in Fig. 1) separately. In \cite{Hu2021}, a dual-link pilot transmission method was presented, which can estimate the $\Hm_{\rm TI}$ and $\Hm_{\rm IR}$ separately with some approximation. However, this method is restricted to the single-antenna transmitter system and an extension to the multiple-antenna case is non-trivial. To the best of our knowledge, it is generally impractical to estimate them separately due to the lack of the RF chains in the IRS.


As investigated in the existing works \cite{He2021, Chung2024, Chen2013, chung2024efficient, lee2024channel}, there are efficient methods to estimate the so-called effective (or cascaded) channel. Focusing on the SU-MIMO systems with {\em multiple} antennas at the RX, which is the system model considered in this paper, the effective channels in the literature are categorized into two types. In \cite{He2021, Chung2024}, the effective channel is defined as the Kronecker product of the TX-IRS and IRS-RX channels. Due to its large-dimension, this effective channel might not be suitable for the joint optimization of the reflection vector and the hybrid beamforming. Very recently in \cite{lee2024channel}, the effective channel is defined in a more compact form (see Section II for details), thereby being more adequate for the joint optimization. Also, the channel estimation method, proposed in  \cite{lee2024channel} by means of a collaborative low-rank approximation, can yield the best estimation accuracy, while having a lower training overhead. Motivated by this, we in this paper study the joint optimization for IRS-aided mmWave MIMO systems, only using the effective channel in \cite{lee2024channel}. Toward this, our major contributions are summarized as follows.
\begin{itemize}
    \item Due to the highly non-convexity, it is intractable to tackle the aforementioned joint optimization directly. We tackle this challenging problem with the following steps. Relaxing the modulus constraint, we define the {\em relaxed} optimization problem whose solution can yield the upper-bound on the maximum spectral efficiency. Thanks to the relaxation, the optimal solution of the relaxed problem can be obtained, which includes the {\em relaxed} reflection vector, the {\em relaxed} analog precoder and combiner, and the digital precoder and combiner. We theoretically prove that in the large-system limit, they become the near-optimal solution of the original problem by satisfying the modulus constraints.

    \item Specifically, in Lemma 2, it is proved that our {\em relaxed} reflection vector, which is simply obtained from the singular value decomposition (SVD) of the effective channel in \cite{lee2024channel}, is the near-optimal solution of the relaxed problem. Here, the gap from the optimal performance becomes negligible as signal-to-noise ratio (SNR) grows. In Lemma 3, it is proved that our relaxed reflection vector, and relaxed analog precoder and combiner can satisfy the modulus constraints in the large-system limit. Putting it all together, we can derive the asymptotically near-optimal reflection vector and the hybrid beamforming matrices (see Theorem 1).

    \item Based on our theoretical analysis, we can construct the practical reflection vector and the hybrid beamforming matrices by projecting the near-optimal ones into the modulus constraints (i.e., just taking the phases of the complex values). Via simulations, we demonstrate that the proposed method can outperform the SOTA method in \cite{9743307} for large-but-finite mmWave SU-MIMO systems. Furthermore, combining with the channel estimation method in \cite{lee2024channel}, it is shown that the proposed method is robust to the channel estimation errors, almost achieving the ideal performances while having a lower training overhead. Due to its attractive performance and lower-complexity, our construction would be a good candidate for mmWave MIMO systems.

\end{itemize}

The remaining part of this paper is organized as follows. In Section II, we describe the IRS-aided SU-MIMO systems and define their wireless channel models. We state the main results of this paper in Section III, by providing the asymptotically near-optimal reflection vector and hybrid beamforming matrices. Theoretical analysis is conducted in Section IV. Section V provides the simulation results and Section VII concludes the paper.

{\em Notations.} Let $[N]\eqdef\{1,2,...,N\}$ for any positive integer $N$. Given a $M \times N$ matrix $\Hm$, let $\Hm(i,:)$ and $\Hm(:,j)$ denote the $i$-th row and $j$-th column of $\Hm$, respectively. Also, given $m < M$ and $n < N$, let $\Hm([m],:)$ and $\Hm(:,[n])$ denote the submatrices by taking the first $m$ rows and $n$ columns of $\Hm$, respectively. We use $\xv$ and $\Hm$ to denote a column vector and matrix, respectively. Also, $\otimes$ denotes the Kronecker product. Given any two same dimension of vectors $\xv$ and $\yv$, let $\xv\circ\yv$ denotes the Hadamard product of $\xv$ and $\mbox{conj}(\yv)$, where $\mbox{conj}(\yv)$ and $\mbox{conj}(\Hm)$ is the complex conjugate vector and matrix of $\yv$ and $\Hm$, respectively. And, given a vector $\vv$, $\mbox{diag}(\vv)$ denotes a diagonal matrix whose $\ell$-th diagonal element is equal to the $\ell$-th element of $\vv$. We let $\Id$ and ${\bf 0}$ denote the identity and all-zero matrices, respectively, where the sizes of these matrices are easily obtained from the context. Given a vector $\vv$ (resp., a matrix $\Vm$), $\Pc(\vv)$ (resp. $\Pc(\Vm)$) represents the {\em projection operator} taking only the phase of each element of $\vv$ (resp. $\Vm$). Without loss of generality, it is assumed that in the diagonal matrix of SVD (or eigen-decomposition), the diagonal elements (i.e., singular values or eigenvalues) are sorted in the descending order of their absolute values.

\begin{figure}[t]
\centering
\includegraphics[width=0.9\linewidth]{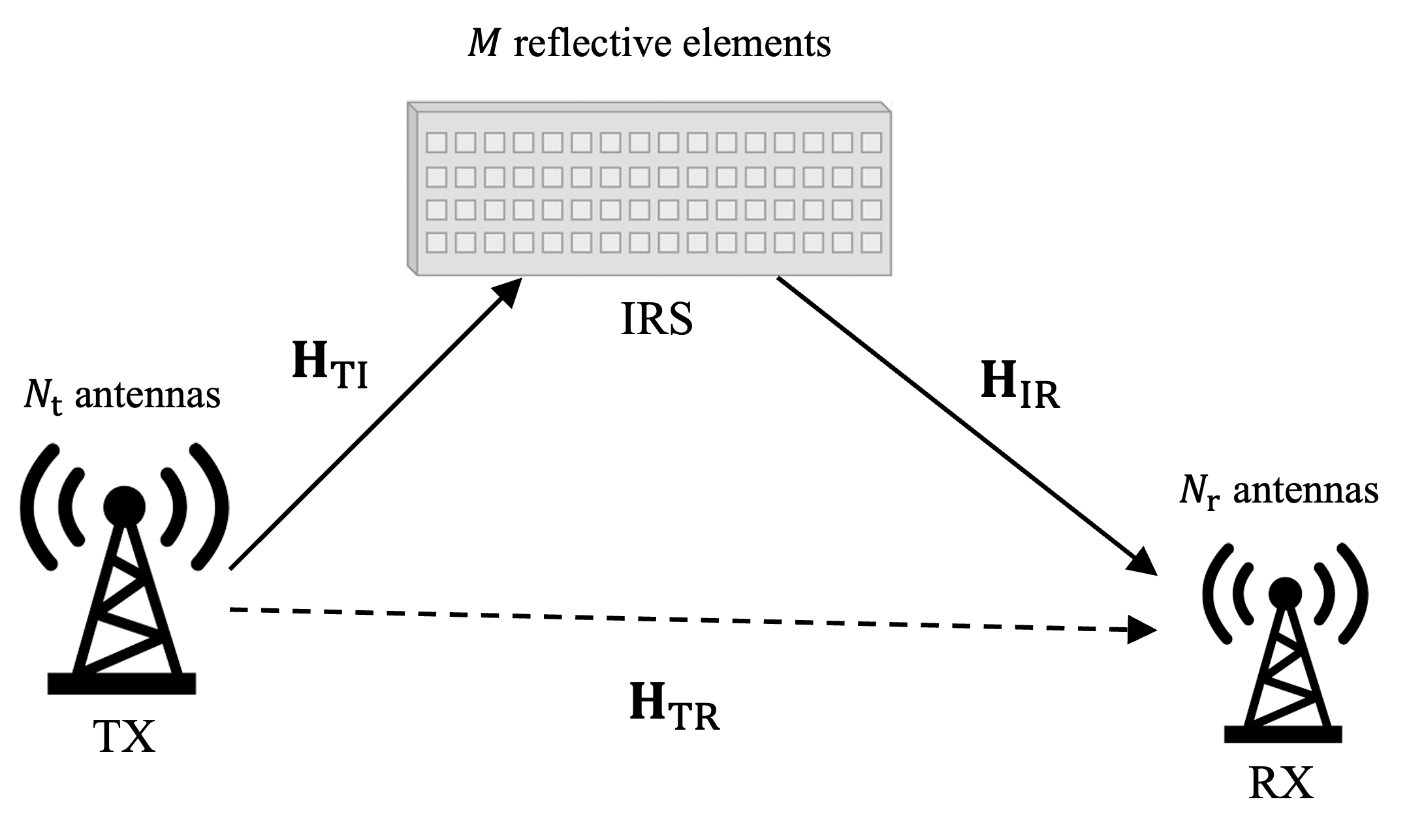}
\caption{Description of an IRS-aided mmWave MIMO system.}
\end{figure}

\section{System Model}\label{sec:System Model}

We consider a narrowband mmWave MIMO system with a low-cost passive intelligent reflecting surface (IRS), which is illustrated in Fig. 1. In this system, the IRS is equipped with $M$ reflective elements to assist the communication between the TX with $N_{\rm t}$ antennas and the RX with $N_{\rm r}$ antennas. For the sake of lower complexity, cost, and power consumption \cite{ahmed2018survey, molisch2017hybrid}, both TX and RX are assumed to employ the hybrid beamforming architecture with a limited number of RF chains, where $N_{\rm t}^{\rm RF}\leq N_{\rm t}$ and $N_{\rm r}^{\rm RF}\leq N_{\rm r}$ denote the number of RF chains at the TX and the RX, respectively. As shown in Fig. 1, the channel responses from the TX to the IRS (i.e., the TX-IRS channel), from the IRS to the RX (i.e., the IRS-RX channel), and from the TX to the RX (i.e., the TX-RX channel) are respectively denoted as $\Hm_{\rm TI} \in \CC^{M\times N_{\rm t}}$, $\Hm_{\rm IR} \in \CC^{N_{\rm r} \times M}$, and $\Hm_{\rm TR} \in \CC^{N_{\rm r}\times N_{\rm t}}$. In the IRS, a reflection vector is defined as
\begin{equation}
    \vv=[v_1,v_2,...,v_M]^{\herm} \; \mbox{with}\; v_{m} = e^{j\vartheta_{m}},
\end{equation} where $\vartheta_{m}\in[0,2\pi)$ is the phase shift of the $m$-th reflective element. Including this, the {\em total} channel response from the TX to the RX is defined as
\begin{equation}
    \Hm_{\rm tot}(\vv)=\Hm_{\rm TR} + \Hm_{\rm IR}\mbox{diag}(\vv)\Hm_{\rm TI}. \label{eq:totchannel}
\end{equation} To optimize the hybrid beamforming from the $\Hm_{\rm tot}(\vv)$, it is required to estimate the channels  $\Hm_{\rm IR}$ and $\Hm_{\rm TI}$ separately. Such optimization has been investigated in \cite{9743307}. However, in practice, it is impractical to estimate them separately due to the lack of RF chains in the IRS. As studied in the existing works \cite{He2021, Chung2024, Chen2013, chung2024efficient, lee2024channel}, there exist low-complexity methods to estimate the so-called effective channel (or the cascaded channel). Following the best-known method in \cite{lee2024channel}, the effective channel is defined as follows. The total channel response in \eqref{eq:totchannel} can be rewritten as
\begin{equation}
    \Hm_{\rm tot}(\vv) =\Hm_{\rm TR} + \Hm_{\rm eff}(\Id \otimes \vv),\label{eq:channel model}
\end{equation} where the (IRS-aided) effective channel is defined as
\begin{align}\label{eff:channels}
    \Hm_{\rm eff}&\eqdef\begin{bmatrix}
        \Hm_{[{\rm eff},1]} & \cdots & \Hm_{[{\rm eff},N_{\rm t}]}
    \end{bmatrix}\in\CC^{N_{\rm r} \times M N_{\rm t}},
\end{align} and $\Hm_{[{\rm eff},t]} =\Hm_{\rm IR}\mbox{diag}\left(\Hm_{\rm TI}(:,t)\right)\in\CC^{N_{\rm r} \times M}$.
Motivated by this, we in this paper aim to jointly optimize the reflection vector and the hybrid beamforming matrices, only exploiting the estimated effective channel $\Hm_{\rm eff}$. Focusing on the joint optimization, in the following sections, it is assumed that $\Hm_{\rm eff}$ (equivalently, $\Hm_{[{\rm eff},t]}$'s) is perfectly estimated. In our simulations, the impact of channel estimation errors will be investigated (see Section V).


\subsection{IRS-aided mmWave MIMO systems}

We describe the wireless channel in the IRS-aided mmWave MIMO system. It is assumed that the TX, the IRS, and the RX are equipped with a uniform planar array (UPA).
Applying the physical propagation model of the wireless channel \cite{tsai2018efficient}, each of the channel responses $\Hm_{i}$ is given by \cite{liu2020matrix}:
\begin{equation}
    \Hm_{i}=\sum_{s=1}^{N_{\rm path}^{i}}\alpha_{[i,s]}\av_{\rm r}(\phi_{[i,s]}^{\rm r}, \theta_{[i,s]}^{\rm r})\av_{\rm t}^{\herm}(\phi_{[i,s]}^{\rm t}, \theta_{[i,s]}^{\rm t}),\label{eq:F}
\end{equation} for $i \in \{{\rm TR},{\rm IR},{\rm TR}\}$,  where $N_{\rm path}^{i}$ is the number of the associated spatial paths. Also, $\alpha_{[i,s]}$ denotes the complex gain of the $s$-th spatial path in the $\Hm_{i}$, which is independently distributed with $\mathcal{CN}(0,\gamma_i^{2}10^{-0.1PL(d_i)})$, where the normalization factor is given as
\begin{equation*}
    \gamma_{i} = \sqrt{{\rm col}\left(\Hm_{i}\right){\rm row}\left(\Hm_{i}\right)/N_{\rm path}^{i}},
\end{equation*} and $PL(d_i)$ is the path loss caused by the distance $d_i$ between the two associated entities \cite{Akden2014}. Here, ${\rm col}(\Hm_i)$ and ${\rm row}(\Hm_i)$ denote the number of columns and rows of $\Hm_{i}$, respectively. In the $s$-th path of $\Hm_{i}$, the normalized receiver UPA steering vector is represented as
\begin{align}\label{eq:sv1}
    &\av_{\rm r}(\phi_{[i,s]}^{\rm r}, \theta_{[i,s]}^{\rm r})\\
    &=\frac{1}{\sqrt{N_{\rm r}}}\left[1,\dots,e^{j\frac{2\pi d}{\lambda}(i_h\sin(\phi_{[i,s]}^{\rm r})\sin(\theta_{[i,s]}^{\rm r})+i_v\cos(\theta_{[i,s]}^{\rm r}))},\right.\nonumber\\
    &\left.\dots,e^{j\frac{2\pi d}{\lambda}((N_{\rm r}^{h}-1)\sin(\phi_{[i,s]}^{\rm r})\sin(\theta_{[i,s]}^{\rm r})+(N_{\rm r}^{v}-1)\cos(\theta_{[i,s]}^{\rm r}))}\right]^{\intercal},\nonumber
\end{align} where $\lambda$ is the signal wavelength and $d$ is the spacing between the antennas or IRS elements. The horizontal and vertical indices for the receive antennas are respectively denoted by $0\leq i_h <N_{\rm r}^{h}$ and $0\leq i_v <N_{\rm r}^{v}$, where $N_{\rm r}=N_{\rm r}^{h}N_{\rm r}^{v}$. Also, the normalized transmit UPA steering vector of the $s$-th path in $\Hm_{i}$ is defined as $\av_{\rm t}(\phi_{[i,s]}^{\rm t},\theta_{[i,s]}^{\rm t})$, which is identically represented as in \eqref{eq:sv1}, by substituting $\phi_{[i,s]}^{\rm r}$ and $\theta_{[i,s]}^{\rm r}$ into $\phi_{[i,s]}^{\rm t}$ and $\theta_{[i,s]}^{\rm t}$, respectively.

\subsection{Hybrid Beamforming}
In the IRS-aided MIMO system with hybrid beamforming, the TX sends $N_{\rm s} \leq \min \{N_{\rm t}^{\rm RF},N_{\rm r}^{\rm RF}\}$  data streams to the RX using the digital precoder $\Fm_{\rm BB}\in\CC^{N_{\rm t}^{\rm RF} \times N_{\rm s}}$ and the analog precoder $\Fm_{\rm RF}\in\CC^{N_{\rm t} \times N_{\rm t}^{\rm RF}}$. On construction, the total transmit power constraint is imposed:
\begin{equation}
    \left\|\Fm_{\rm RF}\Fm_{\rm BB}\right\|_F^2 \leq P_{\rm TX}.
\end{equation} To recover the data streams, the RX uses the analog combiner $\Wm_{\rm RF}\in\CC^{N_{\rm r} \times N_{\rm r}^{\rm RF}}$ and digital combiner $\Wm_{\rm BB}\in\CC^{N_{\rm r}^{\rm RF} \times N_{\rm s}}$. Since the analog beamforming matrices $\Wm_{\rm RF}$ and $\Fm_{\rm RF}$  are implemented phase shifters only, the so-called constant modulus constraint is imposed on each of their elements:
\begin{align}
    |\Wm_{\rm RF}(i,j)|&=1/\sqrt{N_{\rm r}}\\
    |\Fm_{\rm RF}(i,j)|&=1/\sqrt{N_{\rm t}}, 
\end{align} for $\forall (i,j)$. Then, the processed received signal is represented as
\begin{equation}
    \yv = \Wm_{\rm BB}^{\herm}\Wm_{\rm RF}^{\herm}\left(\Hm_{\rm tot}(\vv)\Fm_{\rm RF}\Fm_{\rm BB}\sv + \nv\right),\label{eq:received}
\end{equation} where $\sv\in\CC^{N_{\rm s}\time1}$ is the data symbol vector with $\EE[\sv\sv^{\herm}]=\Id$ and $\nv\in\CC^{N_{\rm r}\times1}$ is an additive white Gaussian noise (AWGN) vector whose entries are independently and identically distributed (i.i.d) with $\mathcal{CN}(0,\sigma_{n}^{2})$.

\section{Main Results}\label{Formulation}

Our goal is to jointly optimize the reflection vector and the beamforming matrices with respect to maximizing the achievable spectral efficiency, only exploiting the effective channel in \eqref{eff:channels}. From the processed received signal in \eqref{eq:received}, the achievable spectral efficiency is derived as 
\begin{align}
&R(\Wm_{\rm RF},\Wm_{\rm BB},\vv,\Fm_{\rm RF},\Fm_{\rm BB}) \label{eq:SingleSpectral}\\
&\eqdef \log_{2}\mbox{det}\left(\Id + \Rm_{\rm n}^{-1}\Wm_{\rm BB}^{\herm}\Wm_{\rm RF}^{\herm}\Hm_{\rm tot}(\vv)\Fm_{\rm RF}\Fm_{\rm BB}\right.\nonumber\\
&\left.\;\;\;\;\;\;\;\;\;\;\;\;\;\;\;\;\;\;\;\;\;\;\;\;\;\;\;\;\;\;\;\;\times\Fm_{\rm BB}^{\herm}\Fm_{\rm RF}^{\herm}\Hm_{\rm tot}(\vv)^{\herm}\Wm_{\rm RF}\Wm_{\rm BB}\right),\nonumber
\end{align} where $\Rm_{\rm n}=\sigma_{\rm n}^{2}\Wm_{\rm BB}^{\herm}\Wm_{\rm RF}^{\herm}\Wm_{\rm RF}\Wm_{\rm BB}$. 
It is extremely challenging to directly solve this optimization due to the non-convexity \cite{el2012capacity, el2014spatially, Heath2016, Yu2016, lin2019hybrid, 9743307}. Instead of directly tackling it, we first consider the {\em relaxed} optimization problem on the tricky modulus constraints, defined as 
\begin{align}
R_{\rm upper}\eqdef &\max R(\Wm_{\rm RF},\Wm_{\rm BB},\vv,\Fm_{\rm RF},\Fm_{\rm BB})\nonumber\\
&\mbox{subject\;to}\;\; \vv^{\herm}\vv=M\nonumber\\
&\;\;\quad\quad\quad\quad \Wm_{\rm RF}^{\herm}\Wm_{\rm RF}=\Id\nonumber\\
&\;\;\quad\quad\quad\quad \Fm_{\rm RF}^{\herm}\Fm_{\rm RF}=\Id\nonumber\\
&\;\quad\quad\quad\quad \left\|\Fm_{\rm RF}\Fm_{\rm BB}\right\|_F^{2}\leq P_{\rm TX}.\label{eq:P2}
\end{align}

The key steps to derive the near-optimal solution to the original problem in \eqref{eq:SingleSpectral} are following:
\begin{itemize}
    \item Thanks to the relaxation, we first derive the optimal solution of the relaxed problem in the large-system limit (see Lemma 1 and Lemma 2).
    \item We then prove that they converge to the near-optimal solution of the original problem by satisfying the modulus constraints (see Lemma 3 and Theorem 1).
    \item Based on our theoretical analysis, we construct the practical reflection vector and hybrid beamforming matrices with an attractive performance.
\end{itemize}

Before stating the main result, we provide the useful definition and derive the key supporting lemmas.
\begin{definition}
    Given a total channel response $\Hm_{\rm tot}(\vv)$, the maximum (achievable) spectral efficiency is defined as
    \begin{align}\label{eq:R_max}
        R_{\rm max}(\Hm_{\rm tot}(\vv))\eqdef \sum_{\ell=1}^{N_s}\log_{2}\left(1+\frac{P_{\ell}}{\sigma_n^2}(\Sigma_{\rm tot}(\ell,\ell))^2 \right),
    \end{align} where $\Hm_{\rm tot}(\vv)=\Um_{\rm tot}\Sigma_{\rm tot}\Vm_{\rm tot}^{\herm}$ via SVD and the optimal power allocations $P_{\ell}$'s are determined by the water-filling: 
    \begin{equation}\label{eq:optimal_power}
    P_{\ell} = \max\left\{\frac{1}{P_{\rm TX}} - \frac{\sigma_n^2}{(\Sigma_{\rm tot}(\ell,\ell))^2}, \;0\right\},\; \ell \in [N_s],
    \end{equation} with $\sum_{\ell=1}^{N_{\rm s}}P_\ell=P_{\rm TX}$. It can be achieved using fully-digital SVD-based MIMO transmission.
    \flushright$\blacksquare$
\end{definition}

\begin{lemma} For any feasible solution in \eqref{eq:P2}, we have the upper-bound:
\begin{equation}
    R(\Wm_{\rm RF},\Wm_{\rm BB},\vv,\Fm_{\rm RF},\Fm_{\rm BB}) \leq R_{\rm max}(\Hm_{\rm tot}(\vv)).\label{eq:Lemma1}
\end{equation} 
Letting $\Hm_{\rm tot}(\vv) = \Um_{\rm tot}\Sigma_{\rm tot}\Vm_{\rm tot}^{\herm}$ via SVD, the upper-bound is achieved with equality if
\begin{align*}
    \Wm_{\rm RF} &= \Um_{\rm tot}(:, [N_{\rm r}^{\rm RF}])\\
    \Wm_{\rm BB} &=\Um_{\rm tot}^{\herm}(:, [N_{\rm r}^{\rm RF}])\Um_{\rm tot}(:, [N_{s}])\\
    \Fm_{\rm RF} &=\Vm_{\rm tot}(:, [N_{\rm t}^{\rm RF}])\\
    \Fm_{\rm BB} &=\Vm_{\rm tot}^{\herm}(:, [N_{\rm t}^{\rm RF}])\Vm_{\rm tot}(:, [N_{s}])\Pm_{\rm WF},
\end{align*} where $\Pm_{\rm WF} =\mbox{diag}([\sqrt{P_1},\dots,\sqrt{P_{N_s}}])$ and $P_{\ell}$'s are given in \eqref{eq:optimal_power}. Namely, these are the optimal solutions of the relaxed problem in \eqref{eq:P2}.
\end{lemma}
\begin{IEEEproof}
    The proof is provided in Section~\ref{subsec:lemma1}.
\end{IEEEproof}
From Lemma 1 and by optimizing $\vv$, $R_{\rm upper}$ in \eqref{eq:P2} can be achieved, i.e.,
\begin{equation}
   R_{\rm upper}=\max_{\vv:\; \vv^{\herm}\vv=M} R_{\rm max}(\Hm_{\rm tot}(\vv)). \label{eq:opt_v}
\end{equation}
This optimization is tricky since as shown in \eqref{eq:R_max}, the reflection vector $\vv$ affects the singular values, which in turn determines the optimal power allocations. To handle this challenge, we focus on the case of high-SNRs, which is reasonable in the IRS-aided massive MIMO system \cite{el2014spatially}. In this case, it is well-known that equal-power allocation is near-optimal. Based on this, the maximum achievable spectral efficiency in \eqref{eq:R_max} can be well-approximated as
\begin{align}
    &R_{\rm max}(\Hm_{\rm tot}(\vv)){\approx} \sum_{\ell=1}^{N_{\rm s}}\log_{2}\left(\frac{P_{\rm TX}}{N_{\rm s}\sigma_n^2}(\Sigma_{\rm tot}(\ell,\ell))^2 \right)\nonumber\\
    &\quad\quad\quad\quad \stackrel{(a)}{\leq} N_{\rm s}\log_{2}\left(\frac{P_{\rm TX}}{N_{\rm s}\sigma_n^2} \right) + \log_{2}\left(\|\Hm_{\rm tot}(\vv)\|_F^2\right),\label{eq:highsnr}
\end{align} where (a) is due to the Jensen's inequality. We resort to seeking a {\em near-optimal} reflection vector such as
\begin{equation}
    \vv^{\star}=\argmax_{\vv:\; \vv^{\herm}\vv=M} \|\Hm_{\rm tot}(\vv)\|_F^2.\label{eq:opt_v_app}
\end{equation}

\begin{lemma} 
In the case of either large-system limit (i.e., $N_{\rm t}, N_{\rm r}, M \rightarrow \infty$ with some fixed ratios) or no direct-link channel, the near-optimal reflection vector is derived as
\begin{equation}
     \vv^{\star} = \sqrt{M}\Um_{\rm eff}(:,1),\label{eq:vv}
\end{equation} where $\sum_{t=1}^{N_{\rm t}}\Hm_{[{\rm eff},t]}^{\herm}\Hm_{[{\rm eff},t]} = \Um_{\rm eff}\Lambda_{\rm eff}\Um_{\rm eff}^{\herm}$ via eigen-decomposition.
\end{lemma}
\begin{IEEEproof}
    The proof is provided in Section~\ref{subsec:lemma2}.
\end{IEEEproof} 
\vspace{0.2cm}

In the large-system limit, using $\vv^{\star}$ in Lemma 2, $R_{\rm upper}$ can be nearly achieved as
\begin{equation}
      \Delta_{\rm gap} \eqdef R_{\rm upper} - R_{\rm max}(\Hm_{\rm tot}(\vv^{\star})).
\end{equation} Note that this gap becomes negligible as SNR (or $P_{\rm TX}$) increases. From Lemma 1 and Lemma 2, we can construct the near-optimal solutions of the relaxed problem in \eqref{eq:P2}:
\begin{align}
    \Wm_{\rm RF}^{\star} &= \Um^{\star}_{\rm tot}(:, [N_{\rm r}^{\rm RF}])\nonumber\\
    \Wm_{\rm BB}^{\star} &=\left(\Um^{\star}_{\rm tot}(:, [N_{\rm r}^{\rm RF}])\right)^{\herm}\Um_{\rm tot}^{\star}(:, [N_{s}])\nonumber\\
    \Fm_{\rm RF}^{\star} &= \Vm_{\rm tot}^{\star}(:, [N_{\rm t}^{\rm RF}])\nonumber\\
    \Fm_{\rm BB}^{\star} &= (\Fm_{\rm RF}^{\star})^{\herm}\Vm_{\rm tot}^{\star}(:, [N_{s}])\Pm_{\rm WF}^{\star}, \label{eq:beam}
\end{align} where $\Hm_{\rm tot}(\vv^{\star})=\Um_{\rm tot}^{\star}\Sigma_{\rm tot}(\Vm_{\rm tot}^{\star})^{\herm}$ via SVD.


\vspace{0.2cm}
\begin{lemma}
    In the large system limit, the reflection vector $\vv^{\star}$ and the hybrid beamforming matrices in \eqref{eq:beam} satisfy the modulus constraints in \eqref{eq:SingleSpectral}. In particular, 
    $\vv^{\star}$ in \eqref{eq:vv} converge to the limit $\vv_{\rm limit}$:
    \begin{equation}
        \vv_{\rm limit} = M \av_{\rm r}^{\rm IR}(\phi_{[{\rm IR},1]},\theta_{[{\rm IR},1]})\circ\av_{\rm t}^{\rm TI}(\phi_{[{\rm TI},1]},\theta_{[{\rm TI},1]}).\label{eq:v_asymp}
    \end{equation}
\end{lemma}
\begin{IEEEproof}
    The proof is provided in Section~\ref{subsec:lemma3}.
\end{IEEEproof}

\vspace{0.1cm}
Based on the key lemmas, we state the main result below:
\begin{theorem}
In the large-system limit, the reflection vector in Lemma 2 and the hybrid beamforming matrices in \eqref{eq:beam} can achieve the spectral efficiency $R_{\rm prop}$:
\begin{equation}
    R_{\rm prop} = R_{\rm achiev} - \Delta_{\rm gap},
\end{equation} where $R_{\rm achiev}$ is the maximum achievable spectral efficiency of the original problem in \eqref{eq:SingleSpectral} under the constant modulus constraints.
\end{theorem}
\begin{IEEEproof}
    Combining Lemma 1, Lemma 2, and Lemma 3, the proof is completed.
\end{IEEEproof}
Since $\Delta_{\rm gap}$ can be negligible as SNR (or $P_{\rm TX}$) increases, the proposed method can guarantee the near-optimality in the large-system limit.

We next focus on the {\em practical} RIS-aided MIMO systems, in which the asymptotic results in Lemma 3 might not hold. On the basis of our asymptotic results, the reflection vector is constructed by only applying the projection operator $\Pc(\cdot)$ to the $\vv^{\star}$ in Lemma 2: 
\begin{equation}
    \hat{\vv}^{\star}=\Pc(\vv^{\star}).\label{eq:IRS_p}
\end{equation} Similarly, from \eqref{eq:beam}, the analog beamforming matrices are constructed as
\begin{align}
    \hat{\Wm}_{\rm RF}^{\star} &= \Pc\left(\Um^{\star}_{\rm tot}(:, [N_{\rm r}^{\rm RF}]) \right),\\
    \hat{\Fm}_{\rm RF}^{\star} &= \Pc\left(\Vm_{\rm tot}^{\star}(:, [N_{\rm t}^{\rm RF}])\right).
\end{align} Finally, to satisfy the transmit-power constraint in \eqref{eq:SingleSpectral}, the normalized digital precoder is constructed as
\begin{equation}
        \hat{\Fm}_{\rm BB}^{\star} = \left(\frac{\sqrt{P_{\rm TX}}}{\|\hat{\Fm}_{\rm RF}^{\star}\Fm_{\rm BB}^{\star}\|_F}\right)\times \Fm_{\rm BB}^{\star}.
\end{equation}

\begin{remark} We compare the proposed reflection vector in \eqref{eq:IRS_p} with that in the state-of-the-art (SOTA) method in \cite{9743307}. In fact, the reflection vector in \cite{9743307} is identical to our limit $\vv_{\rm limit}$ in \eqref{eq:v_asymp}. While both methods have identical reflection vector in the large-system limit, they are completely different in practical large-but-finite IRS-aided MIMO systems. Via experiments in Section~\ref{sec:SR}, it is demonstrated that the proposed reflection vector can outperform the SOTA in \cite{9743307}, especially due to the use of a better reflection vector. Noticeably, the proposed $\hat{\vv}^{\star}$ is indeed practical as it can be constructed only using the effective channel in \eqref{eff:channels}. In contrast, to construct the reflection vector in \cite{9743307} (i.e., $\vv_{\rm limit}$ in \eqref{eq:v_asymp}), the TX-IRS and IRS-RX channels should be estimated. More seriously, it is intractable to recover the UPA steering vectors in \eqref{eq:v_asymp} from these channel matrices. We for the first time derive the near-optimal and practical design of the reflection vector and hybrid beamforming for IRS-aided mmWave MIMO systems.
\end{remark}

\section{Theoretical Analysis}

In this section, we provide the theoretical analysis of the proposed reflection vector and the hybrid beamforming, by providing the proofs of the key lemmas.

\subsection{Upper-Bound: Proof of Lemma 1}\label{subsec:lemma1}

We note that $R_{\rm max}(\Hm_{\rm tot}(\vv))$ in Definition 1 is the maximum achievable spectral efficiency when the fully-digital architecture is considered (i.e., $N_{\rm r}^{\rm RF}=N_{\rm r}$ and $N_{\rm t}^{\rm RF}=N_{\rm t}$). Definitely, this is the upper-bound as the hybrid beamforming cannot achieve a higher spectral efficiency than the fully-digital counterpart. We next derive the optimal beamforming matrices for the optimization in \eqref{eq:P2} by achieving the upper-bound $R_{\rm max}(\Hm_{\rm tot}(\vv))$.

We next prove the equality-part by constructing the optimal beamforming matrices for the relaxed problem in \eqref{eq:P2}. Based on the SVD, we have:
\begin{equation}
    \Hm_{\rm tot}(\vv)=\Um_{\rm tot}\Sigma_{\rm tot}\Vm_{\rm tot}^{\herm}.
\end{equation} Thus, we can find the optimal beamforming matrices (that achieve the upper-bound), by solving the equations subject to the feasible solutions in \eqref{eq:P2}:
\begin{align}
    \Wm_{\rm RF}\Wm_{\rm BB} &= \Um_{\rm tot}(:, [N_{s}]) \in \CC^{N_{r}\times N_{s}}\label{eq:rec}\\
    \Fm_{\rm RF}\Fm_{\rm BB} &= \Vm_{\rm tot}(:, [N_{s}])\Pm_{\rm WF} \in \CC^{N_{t}\times N_{s}},\label{eq:trans}
\end{align} where $\Pm_{\rm WF}=\mbox{diag}([\sqrt{P_1},\dots,\sqrt{P_{N_s}}])$ with the optimal power allocations $P_{\ell}$'s in \eqref{eq:optimal_power}. To solve these equations, we first choose the analog combiner $\Wm_{\rm RF}$ as 
\begin{equation}
    \Wm_{\rm RF} = \Um_{\rm tot}(:, [N_{\rm r}^{\rm RF}]),\label{eq:RF}
\end{equation} which is the feasible solution because
\begin{equation*}
    (\Wm^{\rm RF})^{\herm}\Wm^{\rm RF} = \Um_{\rm tot}(:, [N_{\rm r}^{\rm RF}])^{\herm}\Um_{\rm tot}(:, [N_{\rm r}^{\rm RF}]) = \Id.
\end{equation*} Given the $\Wm_{\rm RF}$, we derive the digital combiner $\Wm_{\rm BB}$ as the solution of the resulting linear equation in \eqref{eq:rec}:
\begin{align}
\Wm_{\rm BB} &= \Wm_{\rm RF}^{\dag}\Um_{\rm tot}(:, [N_{s}])\nonumber\\
&=\Um_{\rm tot}(:, [N_{\rm r}^{\rm RF}])^{\herm}\Um_{\rm tot}(:, [N_{s}]),\label{eq:WBB}
\end{align} since $\Wm_{\rm RF}^{\dag}  =\Um_{\rm tot}^{\star}(:, [N_{\rm r}^{\rm RF}])$. We remark that the least-square solution (using the pseudo-inverse) in \eqref{eq:WBB} gives the exact on because the columns of $\Um_{\rm tot}(:, [N_{s}])$ lie in the column space of $\Wm_{\rm RF}$. Thus, the receive beamforming matrices in \eqref{eq:RF} and \eqref{eq:WBB} are the feasible solutions to satisfy the equations in \eqref{eq:rec}. Exactly following the same arguments, we can derive the transmit beamforming matrices such as
\begin{align}
    \Fm_{\rm RF} &=\Vm_{\rm tot}(:, [N_{\rm t}^{\rm RF}]) \label{eq:DRF}\\
    \Fm_{\rm BB} &=\Vm_{\rm tot}(:, [N_{\rm t}^{\rm RF}])^{\herm}\Vm_{\rm tot}(:, [N_{s}])\Pm_{\rm WF}.
\end{align} Definitely, they are the feasible solutions in \eqref{eq:P2} as the power-constraint is satisfied, i.e.,
\begin{align*}
    \|\Fm_{\rm RF}\Fm_{\rm BB}\|_{F}^{2} &= \|\Vm_{\rm tot}(:,[N_s])\Pm_{\rm WF}\|_{F}^2\\
    &=\|\Pm_{\rm WF}\|_{F}^2 =P_{\rm TX}.
\end{align*} This completes the proof of Lemma 1.

\subsection{Optimal Reflection Vector: Proof of Lemma 2}\label{subsec:lemma2}
For ease of exposition, we define the receive array response matrix $\Am_{\rm r}^{i}$, the complex gain matrix $\Gm_{i}$ and the transmit array response matrix $\Am_{\rm t}^{i}$: for $i\in[{\rm IR},{\rm TI},{\rm TR}]$, 
\begin{align}
    \Am_{\rm r}^{i} &=\begin{bmatrix}
         \av_{\rm r}(\phi_{[i,1]}^{\rm r},\theta_{[i,1]}^{\rm r}) &\cdots& \av_{\rm r}(\phi_{[i,N_{\rm path}^{i}]}^{\rm r},\theta_{[i,N_{\rm path}^{i}]}^{\rm r})
    \end{bmatrix},\nonumber\\
    \Gm_{i} &= \mbox{diag}([\alpha_{[{i,1]}},\dots,\alpha_{[i,N_{\rm path}^{i}]}]),\nonumber\\
    \Am_{\rm t}^{i} &=\begin{bmatrix}
         \av_{\rm t}(\phi_{[i,1]}^{\rm t},\theta_{[i,1]}^{\rm t}) &\cdots& \av_{\rm t}(\phi_{[i,N_{\rm path}^{i}]}^{\rm t},\theta_{[i,N_{\rm path}^{i}]}^{\rm t})
    \end{bmatrix}.\label{eq:GM}
\end{align} Using them, the channel responses in \eqref{eq:F} can be rewritten as
\begin{equation}
    \Hm_{i} = \Am_{\rm r}^{i}\Gm_{i}\left(\Am_{\rm t}^{i}\right)^{\herm},\label{eq:vectorform}
\end{equation} for $i\in[{\rm IR},{\rm TI},{\rm TR}]$. Without loss of generality, we assume that the diagonal elements in $\Gm_{i}$ is sorted as $|\alpha_{[i,m]}|\geq|\alpha_{[i,n]}|,\forall m,n\in[N_{\rm path}^{i}]$ with $m < n$.

First, we will prove that in the large-system limit, $\Hm_{\rm TR}\Hm_{\rm TI}^{\herm}$ converges to the all-zero matrix, where 
\begin{equation}
    \Hm_{\rm TR}\Hm_{\rm TI}^{\herm} = \Am_{\rm r}^{\rm TR}\Gm_{\rm TR}\left(\Am_{\rm t}^{\rm TR}\right)^{\herm}\Am_{\rm t}^{\rm TI}\Gm_{\rm TI}^{\herm}\left(\Am_{\rm r}^{\rm TI}\right)^{\herm}.\label{eq:HTR}
\end{equation} In \cite{Chen2013}, the asymptotic orthogonality of UPA steering vector was proved, i.e., 
\begin{equation}
    \lim_{X\rightarrow\infty}\av_{X}^{\herm}(f_1,g_1)\av_{X}(f_2,g_2)=0, \label{eq:AOUPA}
\end{equation} for $\forall f_1 \neq f_2, g_1\neq g_2 \in \RR$, where $\av_{X}(f,g)\in\CC^{X\times1}$ (called the general form of UPA response vector) is defined as
\begin{align}
    &\av_{X}(f,g) \label{eq:ax}\\
    &=\frac{1}{X}[1,\dots,e^{j\frac{2\pi d}{\lambda}(x_{h}f+x_{v}g)},\dots,e^{j\frac{2\pi d}{\lambda}((X_{h}-1)f+(X_{v}-1)g)}]^{\transp}\nonumber
\end{align} where $0\leq x_h <X_h$ and $0\leq x_v <X_v$, where $X = X_hX_v$. From \eqref{eq:AOUPA}, we can immediately show that
\begin{equation}
    \lim_{N_{\rm t}, N_{\rm r}, M \rightarrow \infty}\left(\Am_{\rm t}^{\rm TR}\right)^{\herm}\Am_{\rm t}^{\rm TI}=\mathbf{0}.\label{eq:conv}
\end{equation} This implies that each element of $\Hm_{\rm TR}\Hm_{\rm TI}^{\herm}$ converges to 0 as $N_{\rm t}$ goes to infinity. In \eqref{eq:conv}, it was shown that this convergence occurs rapidly. Namely, the asymptotic orthogonality is nearly satisfied with the $256$ number of antennas or IRS reflective elements.

We note that the matrix multiplication in \eqref{eq:conv} can rapidly approach to the zero vector $\bf{0}$ as the number of transmit antennas (i.e., $N_{\rm t}$) grows.  Based on this, we show that 
\begin{align}
    &\lim_{N_{\rm t}, N_{\rm r}, M \rightarrow \infty} \Hm_{\rm tot}(\vv)(\Hm_{\rm tot}(\vv))^{\herm} \nonumber\\
    & = \Hm_{\rm TR}\Hm_{\rm TR}^{\herm}+\Hm_{\rm IR}\mbox{diag}(\vv)\Hm_{\rm TI}\Hm_{\rm TI}^{\herm}\mbox{diag}(\vv)^{\herm}\Hm_{\rm IR}^{\herm}.\label{eq:limit}
\end{align} 
From the definition of $R_{\rm max}\left(\Hm_{\rm tot}(\vv)\right)$, the (relaxed) IRS reflection vector is optimized by taking the solution of
\begin{align}
\vv^{\star}=&\argmax_{\vv:\; \vv^{\herm}\vv=M}\left\|\Hm_{\rm tot}(\vv)\right\|_F^2,\nonumber\\
=&\argmax_{\vv:\; \vv^{\herm}\vv=M}\left\|\Hm_{\rm TR} + \Hm_{\rm IR}\mbox{diag}(\vv)\Hm_{\rm TI}\right\|_F^2\nonumber\\
\stackrel{(a)}{=}&\;\argmax_{\vv:\; \vv^{\herm}\vv=M}\left(\left\|\Hm_{\rm TR}\right\|_F^2 + \left\|\Hm_{\rm IR}\mbox{diag}(\vv)\Hm_{\rm TI}\right\|_F^2\right),\nonumber\\
=&\;\argmax_{\vv:\; \vv^{\herm}\vv=M}\left\|\Hm_{\rm IR}\mbox{diag}(\vv)\Hm_{\rm TI}\right\|_F^2,
\nonumber\\
\stackrel{(b)}{=}&\;\argmax_{\vv:\; \vv^{\herm}\vv=M}\;\vv^{\herm}\left(\sum_{t=1}^{N_{\rm t}}\Hm_{[{\rm eff},t]}^{\herm}\Hm_{[{\rm eff},t]}\right)\vv \label{eq:vopt}
\end{align} where (a) follows from \eqref{eq:limit}, and (b) is from the definition of the effective channel in \eqref{eq:channel model} and due to the fact that
\begin{align}
\left\|\Hm_{\rm IR}\mbox{diag}(\vv)\Hm_{\rm TI}\right\|_{F}^{2}&=\left\|\Hm_{\rm eff}\left(\Id\otimes\vv\right)\right\|_{F}^{2}\nonumber\\&=\left\|\begin{bmatrix}
        \Hm_{[{\rm eff},1]}\vv & \cdots & \Hm_{[{\rm eff},N_{\rm t}]}\vv
    \end{bmatrix}\right\|_{F}^{2}\nonumber\\
&=\vv^{\herm}\left(\sum_{t=1}^{N_{\rm t}}\Hm_{[{\rm eff},t]}^{\herm}\Hm_{[{\rm eff},t]}\right)\vv.
\end{align} Given the $\sum_{t=1}^{N_{\rm t}}\Hm_{[{\rm eff},t]}^{\herm}\Hm_{[{\rm eff},t]}=\Um_{\rm eff}\Lambdam_{\rm eff}\Um_{\rm eff}^{\herm}$ via eigen-decomposition, the optimal solution of \eqref{eq:vopt} is simply attained by taking the scalar multiple of the eigenvector corresponding to the largest eigenvalue:
\begin{equation*}
    \vv^{\star}=\sqrt{M}\Um_{\rm eff}(:,1).
\end{equation*} Thus, in the large system limit (i.e., $N_t \rightarrow \infty$), $\vv^{\star}$ is the (relaxed) optimal reflection vector. Obviously, when there is no direct-link channel, $\vv^{\star}$ is also the (relaxed) optimal reflection vector because the optimization in \eqref{eq:vopt} holds. This completes the proof of Lemma 2.

\subsection{Asymptotic Optimality: Proof of Lemma 3}\label{subsec:lemma3}

We prove the asymptotic optimality of the beamforming matrices and the reflection vector in Lemma 2 and Lemma 3, respectively.

First, we will prove that $\vv^{\star}$ in Lemma 2 meets the modulus constraints in the large-system limit. Toward this, $\sum_{t=1}^{N_{\rm t}}\Hm_{[\rm eff,t]}^{\herm}\Hm_{[\rm eff,t]}$ is rewritten as
\begin{align}
    &\sum_{t=1}^{N_{\rm t}}\Hm_{[\rm eff,t]}^{\herm}\Hm_{[\rm eff,t]} = \left(\Hm_{\rm IR}^{\herm}\Hm_{\rm IR}\right)\circ\left(\Hm_{\rm TI}\Hm_{\rm TI}^{\herm}\right),\nonumber\\
    &=\left[\sum_{i=1}^{M}\lambda_{i}^{\rm IR}\ev_{i}^{\rm IR}\left(\ev_{i}^{\rm IR}\right)^{\herm}\right]\circ\left[\sum_{j=1}^{M}\lambda_{j}^{\rm TI}\ev_{j}^{\rm TI}\left(\ev_{j}^{\rm TI}\right)^{\herm}\right],\nonumber\\
    &=\sum_{i=1}^{M}\sum_{j=1}^{M}\lambda_{i}^{\rm IR}\lambda_{j}^{\rm TI}\left(\ev_{i}^{\rm IR}\circ\ev_{j}^{\rm TI}\right)\left(\ev_{i}^{\rm IR}\circ\ev_{j}^{\rm TI}\right)^{\herm},\label{eq:eig}
\end{align} where $\Hm_{\rm IR}^{\herm}\Hm_{\rm IR} = \sum_{i=1}^{M}\lambda_{i}^{\rm IR}\ev_{i}^{\rm IR}\left(\ev_{i}^{\rm IR}\right)^{\herm}$ and  $\Hm_{\rm TI}\Hm_{\rm TI}^{\herm} = \sum_{j=1}^{M}\lambda_{j}^{\rm TI}\ev_{j}^{\rm TI}\left(\ev_{j}^{\rm TI}\right)^{\herm}$ via eigen-decomposition.

In \cite{9743307}, it was proved that in the large system limit, the diagonal elements of $\Gm_i$ in \eqref{eq:GM} converge to the singular values of $\Hm_i$. Due to the asymptotic orthogonality, $\Am_{\rm r}$ and $\Am_{\rm t}$ in \eqref{eq:GM} converge to the eigenvectors of $\Hm_i\Hm_i^{\herm}$ and $\Hm_i^{\herm}\Hm_i$, respectively. In other words,  $\{\ev_{s}^{\rm IR}:s\in[N_{\rm path}^{\rm IR}]\}$ and $\{\ev_{s}^{\rm TI}:s\in[N_{\rm path}^{\rm TI}]\}$ become respectively the UPA steering vectors. Using this fact, $\ev_{i}^{\rm IR}\circ\ev_{j}^{\rm TI}$ can be expressed as
\begin{align}
    \ev_{i}^{\rm IR}\circ\ev_{j}^{\rm TI}&=\av_{\rm r}^{\rm IR}(\phi_{[{\rm IR},i]},\theta_{[{\rm IR},i]})\circ\av_{\rm t}^{\rm TI}(\phi_{[{\rm TI},j]},\theta_{[{\rm TI},j]})\nonumber\\
    &\stackrel{(a)}{=}\av_{M}\left(f_{[i,j]},g_{[i,j]}\right),\label{eq:mam}
\end{align} for $\forall i \in [N_{\rm path}^{\rm IR}]$ and $\forall j\in[N_{\rm path}^{\rm TI}]$, where (a) follows the definition of the general form of UPA response vector in \eqref{eq:ax} and
\begin{align*}
    f_{[i,j]}&={\sin}(\phi_{[{\rm IR},i]}){\sin}(\theta_{[{\rm IR},i]}) - {\sin}(\phi_{[{\rm TI},j]}){\sin}(\theta_{[{\rm TI},j]})\\
    g_{[i,j]}&={\cos}(\theta_{[{\rm IR},i]})-{\cos}(\theta_{[{\rm TI},j]}).
\end{align*} This shows that each $\ev_{i}^{\rm IR}\circ\ev_{j}^{\rm TI}$ can be considered as an UPA steering vector. From the asymptotic orthogonality in 
\eqref{eq:AOUPA}, we can get:
\begin{align}
&\left(\ev_{i}^{\rm IR}\circ\ev_{j}^{\rm TI}\right)^{\herm}\left(\ev_{k}^{\rm IR}\circ\ev_{l}^{\rm TI}\right)\nonumber\\
&\quad\quad =\;\av_{M}^{\herm}\left(f_{[i,j]},g_{[i,j]}\right)\av_{M}\left(f_{[k,l]},g_{[k,l]}\right)=0,\label{eq:neweig}
\end{align} for $\forall i,j,k,l \in [M]$ with $i\neq k$ and $j\neq l$. In the large-system limit, thus, \eqref{eq:eig} becomes the ED of $\sum_{t=1}^{N_{\rm t}}\Hm_{[\rm eff,t]}^{\herm}\Hm_{[\rm eff,t]}$.  Since $\sqrt{M}\left(\ev_{1}^{\rm IR}\circ\ev_{1}^{\rm TI}\right)$ is the eigenvector corresponding to the largest eigenvalue $\frac{1}{M}\lambda_1^{\rm IR}\lambda_1^{\rm TI}$,
the optimal reflection vector $\vv^{\star}$ in Lemma 2 is equal to
\begin{align*}
   \vv^{\star}&=\sqrt{M}\Um_{\rm eff}(:,1) \nonumber\\
   & = M\left(\av_{\rm r}^{\rm IR}(\phi_{[{\rm IR},1]},\theta_{[{\rm IR},1]})\circ\av_{\rm t}^{\rm TI}(\phi_{[{\rm TI},1]},\theta_{[{\rm TI},1]})\right),
\end{align*} where the second equality follows from \eqref{eq:mam}. This implies that in the large-system limit,  $\vv^{\star}$ satisfies the modulus constraints. 

Next, we will prove that in the large-system limit, $\Um^{\star}_{\rm tot}(:[N_{\rm r}^{\rm RF}])$ and $\Vm^{\star}_{\rm tot}(:, [N_{\rm t}^{\rm RF}])$ satisfy the modulus constraints in \eqref{eq:SingleSpectral}. Due to the fact that $\vv^{\star}=\sqrt{M}\Um_{\rm eff}(:,1)$ in Lemma 2, we can get:
\begin{align}
    &\lim_{N_{\rm t},N_{\rm r},M\rightarrow\infty}\Hm_{\rm tot}(\vv^{\star})\Hm_{\rm tot}^{\herm}(\vv^{\star}) \nonumber\\
    &=\sum_{s=1}^{N_{\rm path}^{\rm TR}}|\alpha_{s}^{\rm TR}|^{2}\av_{\rm r}(\phi_{[{\rm TR},s]}^{\rm r}, \theta_{[{\rm TR},s]}^{\rm r})\av_{\rm r}^{\herm}(\phi_{[{\rm TR},s]}^{\rm r}, \theta_{[{\rm TR},s]}^{\rm r})\nonumber\\
    &+ |\alpha_{1}^{\rm TI}|^{2}|\alpha_{1}^{\rm IR}|^{2}\av_{\rm r}(\phi_{[{\rm IR},1]}^{\rm r}, \theta_{[{\rm IR},1]}^{\rm r})\av_{\rm r}^{\herm}(\phi_{[{\rm IR},1]}^{\rm r}, \theta_{[{\rm IR},1]}^{\rm r}),\label{eq:totlimit1}
\end{align} and 
\begin{align}
    &\lim_{N_{\rm t},N_{\rm r},M\rightarrow\infty}\Hm_{\rm tot}^{\herm}(\vv^{\star})\Hm_{\rm tot}(\vv^{\star}) \nonumber\\
    &=\sum_{s=1}^{N_{\rm path}^{\rm TR}}|\alpha_{s}^{\rm TR}|^{2}\av_{\rm t}(\phi_{[{\rm TR},s]}^{\rm t}, \theta_{[{\rm TR},s]}^{\rm t})\av_{\rm t}^{\herm}(\phi_{[{\rm TR},s]}^{\rm t}, \theta_{[{\rm TR},s]}^{\rm t})\nonumber\\
    &+ |\alpha_{1}^{\rm TI}|^{2}|\alpha_{1}^{\rm IR}|^{2}\av_{\rm t}(\phi_{[{\rm TI},1]}^{\rm t}, \theta_{[{\rm TI},1]}^{\rm t})\av_{\rm t}^{\herm}(\phi_{[{\rm TI},1]}^{\rm t}, \theta_{[{\rm TI},1]}^{\rm t}).\label{eq:totlimit2}
\end{align}
Similarly to \eqref{eq:vectorform}, they can be rewritten as
\begin{align*}
    &\lim_{N_{\rm t},N_{\rm r},M\rightarrow\infty}\Hm_{\rm tot}(\vv^{\star})\Hm_{\rm tot}^{\herm}(\vv^{\star}) = \Am_{\rm r}^{\rm tot}\Thetam_{\rm tot}\left(\Am_{\rm r}^{\rm tot}\right)^{\herm}\\
    &\lim_{N_{\rm t},N_{\rm r},M\rightarrow\infty}\Hm_{\rm tot}^{\herm}(\vv^{\star})\Hm_{\rm tot}(\vv^{\star}) = \Am_{\rm t}^{\rm tot}\Thetam_{\rm tot}\left(\Am_{\rm t}^{\rm tot}\right)^{\herm}
\end{align*}
where 
\begin{align}
    \Am_{\rm r}^{\rm tot} &= \begin{bmatrix}
        \Am_{\rm r}^{\rm TR} \;\;\av_{\rm r}(\phi_{[{\rm IR},1]}^{\rm r}, \theta_{[{\rm IR},1]}^{\rm r})
    \end{bmatrix}\in\CC^{N_{\rm r}\times(N_{\rm path}^{\rm TR}+1)}\nonumber\\
    \Thetam_{\rm tot} &= \mbox{diag}\left([|\alpha_{1}^{\rm TR}|^{2},\dots,|\alpha_{N_{\rm path}^{\rm TR}}^{\rm TR}|^{2},|\alpha_{1}^{\rm TI}|^{2}|\alpha_{1}^{\rm IR}|^{2}]\right)\nonumber\\
    \Am_{\rm t}^{\rm tot} &= \begin{bmatrix}
        \Am_{\rm t}^{\rm TR} \;\;\av_{\rm t}(\phi_{[{\rm TI},1]}^{\rm t}, \theta_{[{\rm TI},1]}^{\rm t})
    \end{bmatrix}\in\CC^{N_{\rm t}\times(N_{\rm path}^{\rm TR}+1)}.\label{eq:theta_tot}
\end{align} As proved in the above, the diagonal elements of $\Thetam_{\rm tot}$ asymptotically converge to the eigenvalues of $\Hm_{\rm tot}^{\herm}(\vv^{\star})\Hm_{\rm tot}(\vv^{\star})$. From the asymptotic orthogonality of UPA in \eqref{eq:conv}, the columns of $\Am_{\rm r}^{\rm tot}$ and $\Am_{\rm t}^{\rm tot}$ are orthogonal to each other. Using this fact, we can see that in the large-system limit, $\Um_{\rm tot}^{\star}(:,[N_{\rm path}^{\rm TR}+1])$, $\Sigmam^{\star}_{\rm tot}([N_{\rm path}^{\rm TR}+1],[N_{\rm path}^{\rm TR}+1])$ and $\Vm^{\star}_{\rm tot}(:,[N_{\rm path}^{\rm TR}+1])$ converge to $\Am_{\rm r}^{\rm tot}$, $\Thetam_{\rm tot}$, $\Am_{\rm t}^{\rm tot}$, respectively. Therefore, $ \Um^{\star}_{\rm tot}(:, [N_{\rm r}^{\rm RF}])$ and $ \Vm_{\rm tot}^{\star}(:, [N_{\rm t}^{\rm RF}])$ can satisfy the modulus constraints. This completes the proof of Lemma 3.

\begin{figure}[t]
\centering
\includegraphics[width=1.0\linewidth]{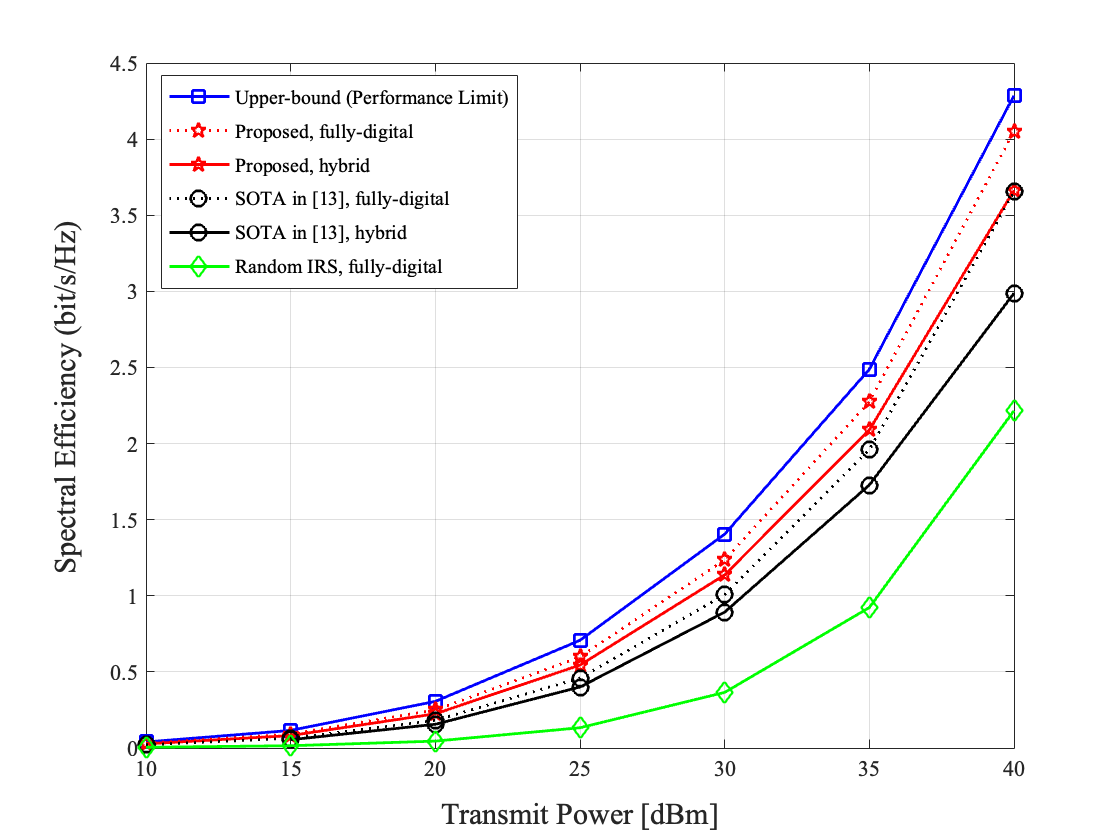}
\caption{The impact of the transmit power on the spectral efficiency. $N_{\rm r} = 4\times4 = 16, N_{\rm t} = 8\times8 = 64, M = 16\times16 = 256$ and $N_{\rm r}^{\rm RF} = N_{\rm t}^{\rm RF} = 4$}
\end{figure}

\begin{figure}[t]
\centering
\includegraphics[width=1.0\linewidth]{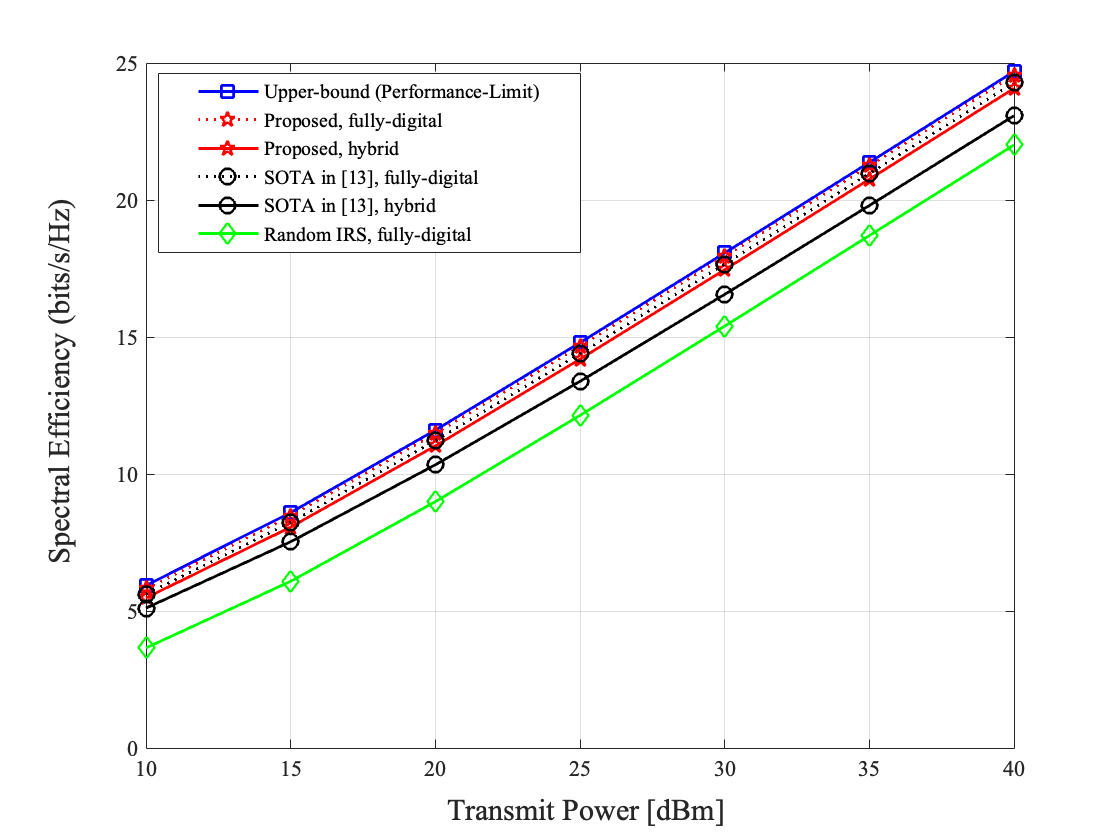}
\caption{The impact of the transmit power on the spectral efficiency. $N_{\rm r} = N_{\rm t} = M = 16\times16 = 256$ and $N_{\rm r}^{\rm RF} = N_{\rm t}^{\rm RF} = 4$.}
\end{figure}
\begin{figure}[t]
\centering
\includegraphics[width=1.0\linewidth]{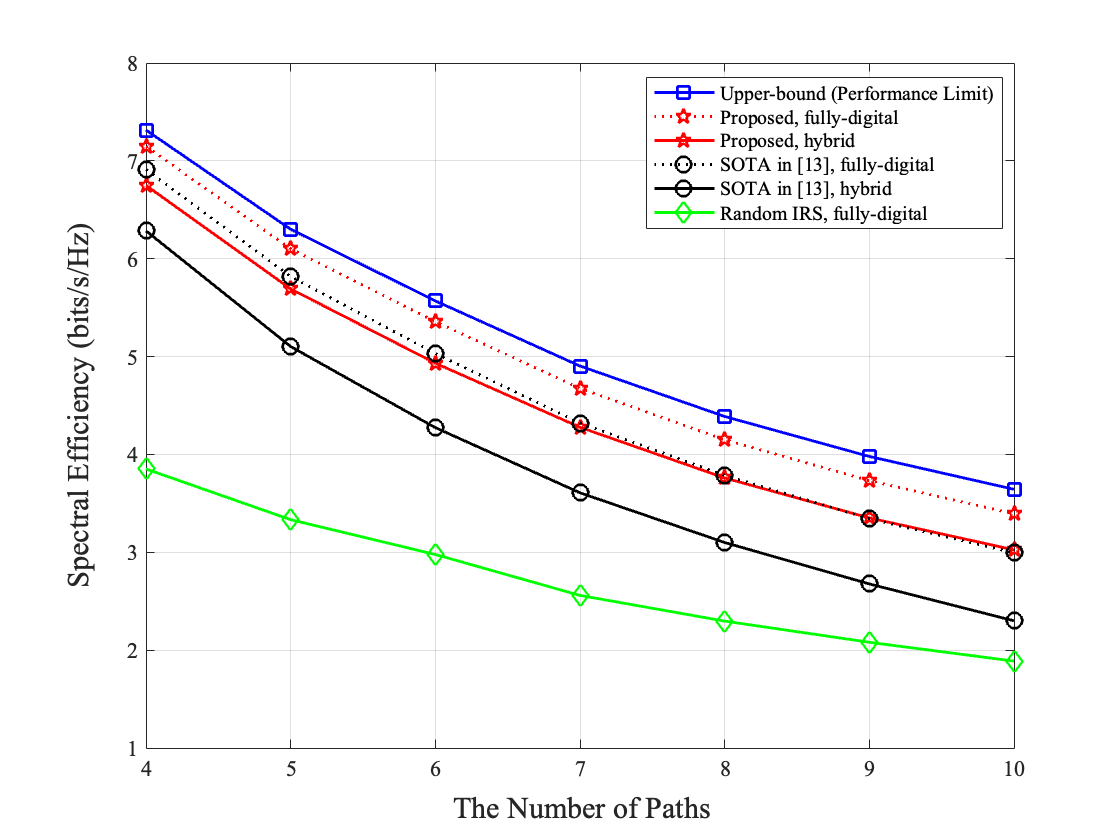}
\caption{The impact of the number of paths on the spectral efficiency. $N_{\rm r} = 4\times4 = 16, N_{\rm t} = 8\times8 = 64, M = 16\times16 = 256$ and $P_{\rm TX} = 40{\rm dBm}$}
\end{figure}
\begin{figure}[t]
\centering
\includegraphics[width=1.0\linewidth]{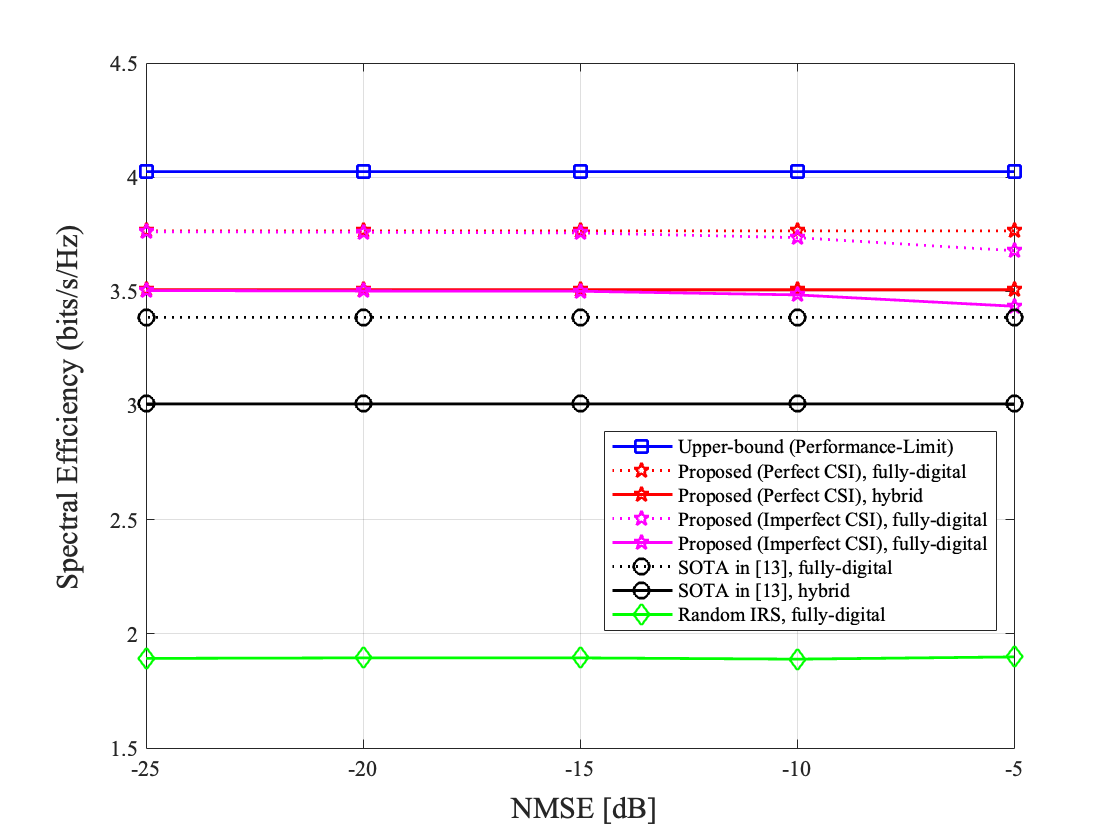}
\caption{The impact of channel estimation accuracy on the spectral efficiency. $N_{\rm r} = 4\times4 = 16, N_{\rm t} = 8\times8 = 64, M = 16\times16 = 256$ and $P_{\rm TX} = 40{\rm dBm}$}
\end{figure}

\section{Simulation Result}\label{sec:SR}
We consider the IRS-aided mmWave MIMO systems in Fig. 1. 
The number of paths are set by $N_{\rm path}^{\rm TR}=N_{\rm path}^{\rm IR}=N_{\rm path}^{\rm TI} = 8$.
Regarding the distances among the TX, IRS, and RX, $d_{\rm TR}$ follows a uniform distribution over the range $\{d_{\rm TI}+d_{\rm IR}-10\;{\rm m},d_{\rm TI}+d_{\rm IR}\;{\rm m}\}$, where $d_{\rm TI}$ and $d_{\rm IR}$ are uniformly distributed over $[50\;{\rm m}, 60\;{\rm m}]$ and $[10\;{\rm m}, 20\;{\rm m}]$, respectively.
Then, the distance-dependent path loss $PL(d_i)$ is modeled as
\begin{equation}
    PL(d_i)\;[{\rm dB}] = \alpha + 10\beta\log_{10}(d_i) + \epsilon,\label{eq:pathloss}
\end{equation} for $i \in [{\rm TR}, {\rm TI}, {\rm IR}]$, where $\epsilon \sim \mathcal{N}(0,\sigma^{2})$. According to the experimental data for 28GHz channels in \cite{Akden2014}, the parameters in \eqref{eq:pathloss} are set by $\alpha = 61.4$, $\beta = 2$, $\sigma=5.8\;{\rm dB}$ for the line-of-sight (LOS) path of $\Hm_i$, and $\alpha = 72.0$, $\beta = 2.92$, $\sigma=8.7\;{\rm dB}$ for the non-line-of-sight (NLOS) paths. To evaluate the effectiveness of various IRS vector designs more accurately, it is assumed that each path of $\Hm_{\rm TR}$ is a NLOS path that passed through tinted-glass walls to experience an additional penetration loss of 40.1 $\rm dB$ \cite{Zhao2013}. The element spacing, noise power, and the number of data streams are each set by $d=\lambda/2$, $\sigma_{n}^{2}=-91{\rm dBm}$, and $N_{\rm s}=4$. Lastly, all the simulation results are averaged over $10^4$ channel realizations.


Fig. 2 shows the impact of the transmit power $P_{\rm TX}$ on the spectral efficiency. Noticeably, the proposed method can outperform the SOTA in \cite{9743307}, while both can achieve near-optimal performance in the large-system limit. The performance gain is due to the construction of a better reflection vector. Namely, our reflection vector $\hat{\vv}^{\star}$ in \eqref{eq:IRS_p} makes $R_{\rm max}\left(\Hm_{\rm tot}(\vv)\right)$ in Definition 1 higher than the reflection vector in  \cite{9743307}. Comparing the performance of fully-digital beamforming using random reflection vectors, Both the proposed and SOTA methods can attain significant gains, manifesting the effect of the optimized reflection vectors. Also, in comparison between the upper-bound (i.e., the performance-limit) and the proposed fully-digital beamforming, we can identify that the effect of the mismatch between $\vv^{\star}$ in Lemma 2 and its projection $\hat{\vv}^{\star}=\Pc(\vv^{\star})$ is not large. This implies that $\hat{\vv}^{\star}$ is an almost optimal practical reflection vector while ensuring the asymptotic optimality.

Fig. 3 shows the achievable spectral efficiency of the proposed method in a larger MIMO system in which the asymptotic orthogonality is nearly achieved. First, we can identify the asymptotic optimality of the proposed method by comparing the performances in Fig. 2 and Fig. 3, because the achievable spectral efficiency of the proposed method becomes closer to the upper-bound as the size of the system grows. As expected, the achievable spectral efficiency by itself can increase as the size of the system scales up. This can be explained from our theoretical analysis. It is obvious that $R_{\rm upper}$ in \eqref{eq:P2} is determined by the top-$N_{\rm s}$ eigenvalues of the total channel $\Hm_{\rm tot}(\vv^{\star})$. Also, our analysis shows that in the large-system limit, these eigenvalues converge to the diagonal elements of $\Thetam_{\rm tot}$ in \eqref{eq:theta_tot}. Among the diagonal elements, $|\alpha_1^{\rm TI}|^{2}|\alpha_1^{\rm IR}|^{2}$ is the dominant value to determine the $R_{\rm upper}$ under the reasonable assumption that the path loss of the TX-RX channel (or direct channel) is much greater than the others. Also, this product can be computed as
\begin{equation}
    \EE\left[|\alpha_1^{\rm TI}|^{2}|\alpha_1^{\rm IR}|^{2}\right] \geq M^{2}n,
\end{equation}where $n = N_{\rm r}N_{\rm t}10^{-0.1(PL(d_{\rm TI})+PL(d_{\rm IR}))}/N_{\rm path}^{\rm TI}N_{\rm path}^{\rm IR}$. This clearly shows that the achievable spectral efficiency of the proposed method tends to be higher as the system scales up.

Fig. 4 shows the impact of the number of spatial paths on the spectral efficiency. As expected, the achievable spectral efficiency tends to be higher as the number of spatial paths decreases. This shows that the proposed IRS-aided hybrid beamforming is more suitable for mmWave MIMO systems, which are usually composed of low-scattering and a small number of spatial paths \cite{poddar2023tutorial}. Also, as in Fig. 2 and Fig. 3, the proposed method can outperform the SOTA in \cite{9743307} regardless of the number of signal paths (i.e., the channel sparsity).

Fig. 5 shows the impact of the channel estimation accuracy on the spectral efficiency. As in \cite{He2021, Chung2024, Chen2013, chung2024efficient, lee2024channel}, the estimation accuracy is commonly measured by the normalized mean square error (NMSE), given by
\begin{equation}
    {\rm NMSE} \eqdef \frac{\|\Hm_{\rm eff}-\hat{\Hm}_{\rm eff}\|_F^{2}}{\|\Hm_{\rm eff}\|_F^{2}},
\end{equation} where $\Hm_{\rm eff}$ and $\hat{\Hm}_{\rm eff}$ denote the true and estimated effective channel, respectively. We observe that the proposed method can guarantee the robustness of the channel estimation errors. Remarkably, the proposed method with imperfect CSI (i.e., the estimated effective channel) can outperform the SOTA with perfect CSI. Note that we could not evaluate the performance of the SOTA with imperfect CSI since there is no practical channel estimation method. Therefore, we can conclude that the proposed method would be a good practical candidate for IRS-aided mmWave MIMO systems with hybrid architectures.




\section{Conclusion}\label{sec:C}

We investigated the joint optimization of the IRS reflection vector and the hybrid beamforming using the so-called effective channel only. This channel can be estimated via many efficient practical methods based on compressed sensing or low-rank matrix approximation. For the first time, we derived the near-optimal construction in the large-system limit. Imposing the modulus constraints on the asymptotically near-optimal construction, we developed the practical ones. Via experiments, it was demonstrated that the proposed method can outperform the SOTA method while providing more practical merits. Our on-going work is to extend the proposed construction in the wideband OFDM MIMO and the multi-user MIMO systems.


\end{document}